\documentclass[floats,twocolumn,pra]{revtex4-2}
\usepackage{eurosym}
\usepackage{graphics}
\usepackage[thmmarks]{ntheorem}
\usepackage{graphicx}
\usepackage{algorithm}
\usepackage{algpseudocode}
\usepackage{amsmath}
\usepackage{amsfonts}
\usepackage{amssymb}
\usepackage{float}
\usepackage{longtable}
\usepackage{flushend}
\usepackage{epsfig}
\usepackage{latexsym}
\usepackage{theorem}
\usepackage{bbm}
\usepackage{bm}
\usepackage{psfrag}
\usepackage[utf8]{inputenc}
\usepackage[T1]{fontenc}
\usepackage{physics}%
\usepackage{color}
\usepackage{multirow}
\usepackage{float}
\usepackage{mathtools}
\usepackage{soul}
\usepackage{hyperref}
\usepackage{mathrsfs}
\usepackage[mathscr]{eucal}
\usepackage{ulem}
\usepackage[makeroom]{cancel}

\newcommand{\SP}[1]{{{\textcolor{black}{#1}}}}

\newcommand{\mr}[1]{{{\mathrm{#1}}}}

\begin{document}

\title{Fundamental Limits on Polarization Entanglement Distribution in Optical Fiber}
\author{Stefano Pirandola}
\affiliation{Department of Computer Science, University of York, York YO10 5GH, United Kingdom}

\begin{abstract} 
Characterizing the ultimate rates of entanglement distribution is essential for both foundational research and the practical deployment of quantum technologies. To investigate these limits, we introduce an erasure-Pauli channel model describing the distribution of polarization entanglement in optical fiber. For this channel, we derive bounds on the rates of entanglement distribution and related quantum resources under optimal local operations and two-way classical communication (two-way assisted capacities). This framework allows us to determine the optimal repeaterless performance achievable over realistic optical fibers affected by polarization mode dispersion, thereby providing a rigorous benchmark for long-distance polarization-based quantum communication. Finally, we show that both our model and capacity bounds remain robust under the inclusion of detector dark counts.
\end{abstract}

\maketitle

\section{Introduction}

The distribution of entanglement is a fundamental primitive for a wide range of quantum technologies, including quantum communication and security~\cite{BB84,Ekert,six,QKDrev0,QKDrev1}, quantum sensing~\cite{sensing1,sensing2}, and distributed quantum computing~\cite{QC1,QC2,Pirandola2019,Cirac1999,Monroe2014,Kimble2008}. Among the various physical implementations, polarization qubits transmitted through optical fiber represent one of the most practical and scalable platforms for long-distance entanglement distribution~\cite{Treiber2009,Wengerowsky2019}. Here, an important open question is to determine the ultimate rates at which polarization-based entanglement can be distributed when the dominant decoherence mechanisms introduced by the optical fiber and the detection system are taken into account. This is the aim of this work.

We provide a general model for this type of distribution, that we call ``erasure-Pauli'' channel. This is defined as a flagged mixture of photon loss and a Pauli polarization channel. While erasure channels and Pauli channels have been extensively studied~\cite{QC1}, their explicit combination in this form does not appear to have been previously analyzed in full generality\SP{~\cite{Enote}}. For this channel, we bound the ultimate rates for distributing entanglement, transmitting quantum information and generating secret keys. In the specific case where the model reduces to an erasure-dephasing channel, these bounds provide the exact formula of the two-way assisted capacities.    

We then apply our formulas to establish the ultimate rates for distributing polarization in optical fiber, where Alice prepares a Bell pair in polarization, keeps one optical qubit and sends the other to Bob through a standard optical fiber.
In this setting, polarization mode dispersion (PMD) is the main cause of decoherence~\cite{WaiMenyuk1996,AgrawalFOCS,PooleWagner1986}. In this setting, we identify two main regimes: one which is depolarizing-dominated, strongly limited to short distances, and another one which is dephasing-dominated, enabled by active polarization control and able to achieve much longer distances. For the latter, we study the performance of the two-way entanglement distribution capacity and show that this ultimate rate is very robust to the presence of dark counts in the receiving setup.

\section{Erasure-Pauli Channel}
\begin{figure}[t]
\vspace{0.0cm}
\centering
\includegraphics[width=0.45\textwidth]{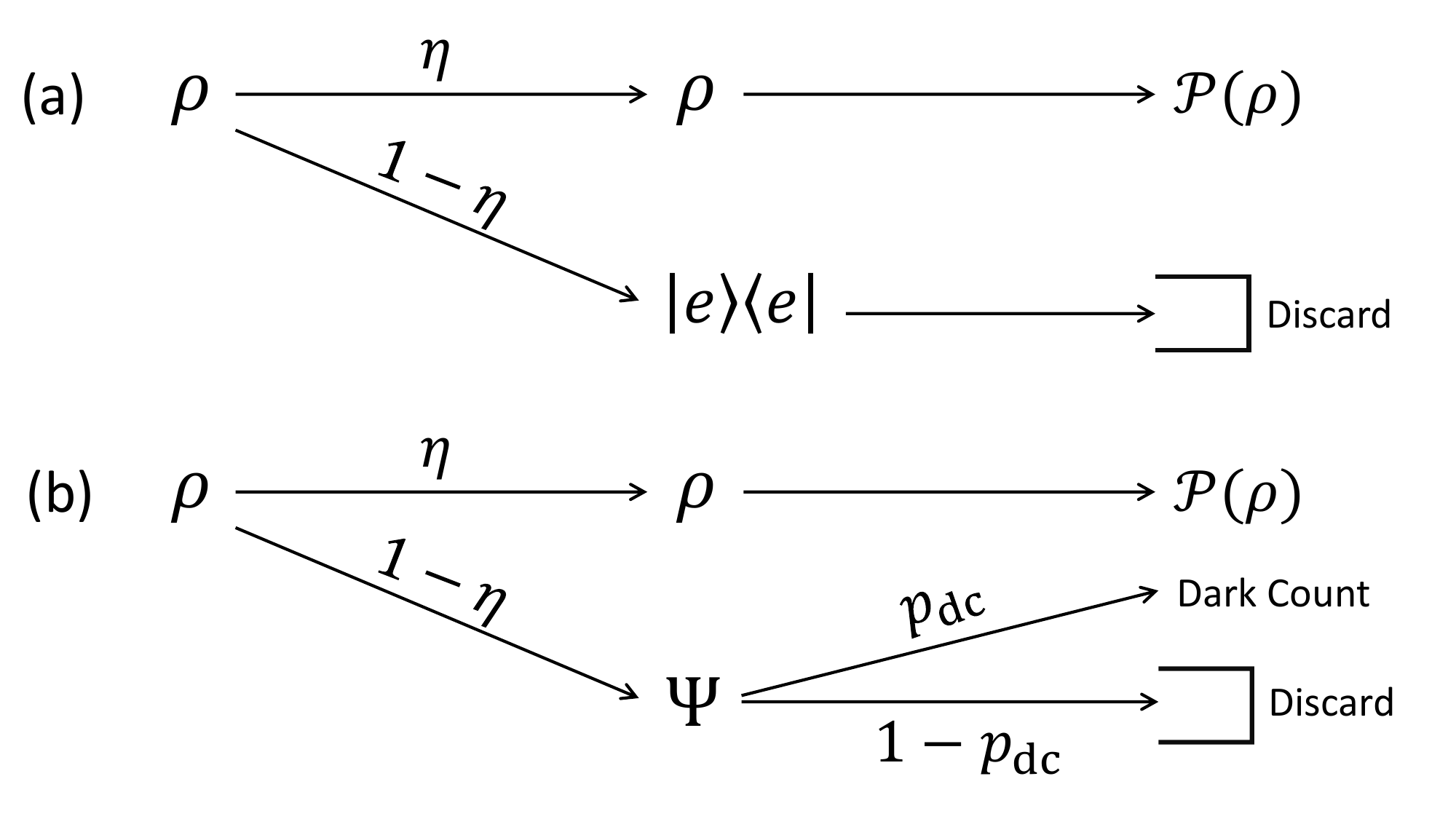}
\caption{Schematic of an erasure-Pauli channel in panel (a) and an erasure-Pauli channel with dark counts in panel (b).}
\label{fig:channel}
\end{figure}

Let us discuss a general type of quantum channel which models loss and noise in fiber-based distribution of entanglement with polarization qubits. Here we need to account for two main effects: (i) the loss of the optical carrier (with a signal or loss event detectable by the receiver); and (ii) the errors induced on the polarization degrees of freedom of the surviving carriers. This process can be modelled by an erasure-Pauli channel. In this general model, the state $\rho$ of an optical polarization qubit undergoes the erasure-Pauli transformation
\begin{equation}
    \mathcal{E}_{\mr{EP}}(\rho)=(1-\eta)|e \rangle \langle e|  +\eta \mathcal{P}(\rho),\label{EPdef}
\end{equation}
where $\eta$ is the transmissivity (probability that the photon is received), $|e \rangle$ is an orthogonal vacuum state (flagging a no detection event at the receiver), and $\mathcal{P}$ is a Pauli channel acting on the polarization of the transmitted qubits (see Fig.~\ref{fig:channel}). Recall that the Kraus representation of a qubit Pauli channel $\mathcal{P}$ can be written as~\cite{QC1}
\begin{align}
\mathcal{P}(\rho)
&=\sum_{k=0}^{3} p_k P_k \rho P_k^\dagger, 
\end{align}
where $\mathbf{p}:=\{p_k\}$ is a probability distribution over the Pauli errors $P_k\in\{I,X,Y,Z\}$, with $I$ the $2 \times2$ identity. 

\section{Capacity bounds}
Let us bound the ultimate rates for entanglement distribution, quantum information transmission, and secret-key agreement over an erasure-Pauli channel with generic transmissivity $\eta$ and error distribution $\mathbf{p}$. These rates are provided by the various two-way assisted capacities, i.e., the two-way assisted entanglement distribution capacity $D_2$, the two-way assisted quantum capacity $Q_2$, and the two-way assisted secret-key capacity $K_2$~\cite{PLOB,2wayCAP}. For simplicity, let us denote by $\mathcal{C}_2$ the generic two-way assisted capacity, which could represent $D_2$, $Q_2$ or $K_2$. While our capacity bounds apply generally to all the considered communication tasks, in the following sections we focus on their application to entanglement distribution over optical fiber.

For an erasure-Pauli channel $\mathcal{E}_{\mr{EP}}$, we show the bounds
\begin{equation}
  \eta [1-H(\mathbf{p})] \le \mathcal{C}_2(\mathcal{E}_{\mr{EP}}) \le \eta \Phi,  \label{EPbounds}
\end{equation}
where $H(\mathbf{p})$ is the Shannon entropy over the Pauli error distribution $\mathbf{p}:=\{p_k\}$, and
\begin{equation}
    \Phi :=  \begin{cases}
1-H_{2}(p_{\mathrm{max}}) & \text{if }p_{\mathrm{max}}\geq\frac{1}{2}\\
0 & \text{otherwise,}
\end{cases} \label{FIg}
\end{equation}
where $H_2$ is the binary Shannon entropy computed over the largest Pauli error $p_{\mathrm{max}}:=\max_k\{ p_k \}$.

\textbf{Proof}.~For the direct part (lower bound), we note that the parties may implement the following entanglement distribution strategy based on backward classical communication. Assume that Alice sends a total of $N$ systems to Bob. When Bob receives the flag state $|e \rangle \langle e|$, Bob discards the system otherwise, he keeps it. This operation can be implemented by Bob applying a dichotomic POVM $\{|e\rangle\langle
e|,I-|e\rangle\langle e|\}$ on each received system and informing Alice which instances were correctly received. On the remaining $\eta N$ systems, each affected by the Pauli channel $\mathcal{P}$, Alice and Bob perform an optimal two-way assisted entanglement distillation protocol. For large $N$, they can distil $D_2(\mathcal{P})$ ebits per surviving system. Therefore, $D_2(\mathcal{E}_{\mr{EP}}) \ge \eta D_2(\mathcal{P})$. From Ref.~\cite{PLOB}, we know that $D_2(\mathcal{P})\geq 1-H(\mathbf{p})$, where $H(\mathbf{p})$ is the Shannon entropy over the Pauli probability distribution. Therefore,
\begin{equation}
    K_2(\mathcal{E}_{\mr{EP}}) \geq Q_2(\mathcal{E}_{\mr{EP}})=D_2(\mathcal{E}_{\mr{EP}}) \geq \eta [ 1-H(\mathbf{p})].
\end{equation}

For the converse part (upper bound), we note that we can rewrite the erasure-Pauli channel as a channel ensemble $\{ \pi_i,\mathcal{E}_i \}$, so that the state undergoes the Pauli channel $\mathcal{E}_0=\mathcal{P}$ with probability $\pi_0=\eta$, and a complete erasure (CE) channel $\mathcal{E}_1=\mathcal{E}_{\mr{CE}}$ with probability $\pi_1=1-\eta$. This is $\mathcal{E}_{\mr{CE}}(\rho)=| e \rangle\langle e|$ for any input state $\rho$. Because these two channels are teleportation-stretchable with resource Choi states $\sigma_i$, we can write the following upper bound for the two-way assisted capacity of the ensemble~\cite{PLOB,condsimulation}
\begin{equation}
    \mathcal{C}_2(\{ \pi_i,\mathcal{E}_i \}) \leq\sum_{i}\pi_{i}E_{\mathrm{R}}
(\sigma_{i})~,
\end{equation}
where $E_R$ is the relative entropy of entanglement~\cite{REE1,REE2,SimeoneBook}. Because, for the CE channel we have $E_R(\sigma_1)=0$ (entanglement breaking) and for the Pauli cannel we have $E_R (\sigma_0)=\Phi$ as in Eq.~\eqref{FIg} (see also Ref.~\cite{PLOB}), we find
\begin{equation}
    \mathcal{C}_2(\mathcal{E}_{\mr{EP}})=\mathcal{C}_2(\{ \pi_i,\mathcal{E}_i \}) \le \eta \Phi.~\blacksquare
\end{equation}

The bounds in Eq.~\eqref{EPbounds} can be simplified for specific distributions of the Pauli errors. In fact, for isotropic errors $\mathbf{p}=\left\{1-3p/4,p/4,p/4,p/4\right\}$, we have an erasure-depolarizing channel $\mathcal{E}_{\mr{EDP}}$ with probability $p$. This channel takes the form as in Eq.~\eqref{EPdef} but where $\mathcal{P}$ simplifies to $\mathcal{P}(\rho)
=(1-p)\rho+pI/2.$ Then, for $Z$-errors only $\mathbf{p}=\left\{1-p,0,0,p\right\}$, we have an erasure-dephasing channel $\mathcal{E}_{\mr{EDH}}$ with probability $p$. This is given by Eq.~\eqref{EPdef} with $\mathcal{P}$ reducing to $\mathcal{P}(\rho)=(1-p)\rho+pZ\rho Z$. For these specific channels, we easily derive~\cite{LBe-dep}
\begin{align}
    &\mathcal{C}_2(\mathcal{E}_{\mr{EDP}}) \leq
    \begin{cases}
 \eta[1-H_{2}\left(  3p/4\right)] & \text{if }p \leq 2/3,  \\
0 & \text{otherwise,}
\end{cases} \label{EDP}
\\
&\mathcal{C}_2(\mathcal{E}_{\mr{EDH}}) =\eta[1-H_2(p)]. \label{EDH}
\end{align}

It is important to identify the necessary and sufficient condition for which the two-way assisted capacity is zero. 
For an erasure-Pauli channel, we may write
\begin{equation}
\mathcal{C}_2(\mathcal{E}_{\mr{EP}}) = 0 \iff \eta=0~~\mr{OR}~~p_{\mathrm{max}} \leq 1/2.\label{EPthreshold}
\end{equation}
For the case of an erasure-depolarizing channel, we therefore have $\mathcal{C}_2(\mathcal{E}_{\mr{EDP}}) = 0$ if and only if $\eta=0$ or $p \geq 2/3$. For the erasure-dephasing,  $\mathcal{C}_2(\mathcal{E}_{\mr{EDH}}) = 0$ if and only if $\eta=0$ (no transmission) or $p = 1/2$ (complete dephasing).

The proof of Eq.~\eqref{EPthreshold} is easy. In fact, the necessary condition ($\impliedby$) trivially comes from Eqs.~\eqref{EPbounds} and~\eqref{FIg}. For the sufficient condition ($\implies$), note that the Choi state of the erasure-Pauli channel is \begin{equation}
\sigma_{\mathcal{E}_{\mr{EP}}}=(1-\eta)\tau_e+\eta \sigma_{\mathcal{P}},
\end{equation}
where $\tau_e=(I/2)\otimes |e \rangle \langle e|$ and $\sigma_{\mathcal{P}}$ is the Choi state of the Pauli channel. The state $\sigma_{\mathcal{E}_{\mr{EP}}}$ is Alice and Bob's joint output state when Alice transmits one qubit of a Bell pair. Assuming that Bob performs the POVM $\{|e\rangle\langle
e|,I-|e\rangle\langle e|\}$ on his qubit, he will project on $\sigma_{\mathcal{P}}$ with probability $\eta$. Conditionally on successful projection, they will then share distillable entanglement $E_D(\sigma_{\mathcal{P}})>0$ for $p_{\mr{max}}>1/2$. In fact, the Choi state of a Pauli channel has non-positive partial transposition~\cite{Peres} for $p_{\max} > 1/2$ and this is equivalent to distillability in dimension $2 \times 2$. 
Therefore, on average, the parties will share $\eta E_D(\sigma_{\mathcal{P}})>0$ ebits. So if $\eta>0$ and $p_{\mr{max}}>1/2$, we have $\mathcal{C}_2(\mathcal{E}_{\mr{EP}}) \geq \eta E_D(\sigma_{\mathcal{P}})>0$. This statement is equivalent to showing the direct implication in Eq.~\eqref{EPthreshold}.   

\section{Polarization noise modelling}
The dominant noise mechanism affecting polarization-based entanglement distribution in optical fiber is PMD. This originates from random birefringence caused by structural imperfections and environmental fluctuations along the fiber. It induces stochastic unitary polarization rotations and, for photons with finite spectral bandwidth, leads to effective phase decoherence due to differential group delay between polarization modes~\cite{AgrawalFOCS}.

In general, this noise mechanism can be modelled as a Pauli channel acting on the polarization degrees of freedom, with a probability distribution that depends on the length and features of the fiber. Including the loss mechanism associated with the propagation of the carrier, the overall channel is therefore an erasure-Pauli channel.  

The basic type of PMD is a depolarizing-dominated Pauli model, with an isotropic error distribution in $X$, $Z$, and $Y$. This regime corresponds to considering an overall erasure-depolarizing channel, where the transmissivity is given by the usual formula $\eta(d) = 10^{-\alpha d/10}$ where $d$ is the fiber-distance in km and $\alpha$ is the loss-rate in dB/km. The depolarizing probability can be expressed as~\cite{PMDnote1}
\begin{equation}
    p(d)=p_{\infty} \left( 1-e^{-d/L} \right), \label{PMDerror}
\end{equation}
where $p_{\infty}$ is the floor value (equal to $1$ for complete depolarization), and $L$ is the polarization decoherence length.
Distribution is not possible for $p(d) \ge 2/3$, corresponding to a maximum fiber-distance of $d_{\mr{max}}=L \ln3$. In the depolarizing-dominated Pauli model, $L$ has a characteristic scale of $10-100$~m, so the range for entanglement distribution is very limited. 

Luckily, the effect of PMD can be mitigated with active polarization control~\cite{PMD1}. Active polarization control suppresses the random unitary rotations, leaving PMD-induced phase decoherence as the dominant mechanism. This is a dephasing-dominated Pauli model which is characterized by a much larger value of the decoherence length $L=L_{\mr{DH}}$. We can write~\cite{PMDnote2}
\begin{equation}
    L_{\mr{DH}}= \frac{2 \tau^2}{D_{\mr{PMD}}^2},\label{PMDcoh}
\end{equation}
where $\tau \simeq (2 \pi \Delta \nu)^{-1}$ is the photon coherence time (with $\Delta \nu$ being the bandwidth) and $D_{\mr{PMD}}$ is the PMD coefficient in ps/$\sqrt{\mr{km}}$. In telecom fibers $D_{\mr{PMD}}\simeq 0.01-10$~ps/$\sqrt{\mr{km}}$~\cite{AgrawalFOCS}. So, for a bandwidth $\Delta \nu=100$~GHz ($\tau \simeq 1.6$~ps), we have a range $L_{\mr{DH}} \simeq 0.05-5 \times 10^4~\mr{km}$. The dephasing probability can be written as in Eq.~\eqref{PMDerror} but with the different floor value $p_{\infty}=1/2$ (equivalent to complete dephasing). In principle, this model would allow for entanglement distribution at any distance, since the threshold $p(d) = 1/2$ corresponds to $L_{\mr{DH}}^{\mr{max}}=\infty$.

\begin{figure}[t]
\vspace{0.2cm}
\centering
\includegraphics[width=0.40\textwidth]{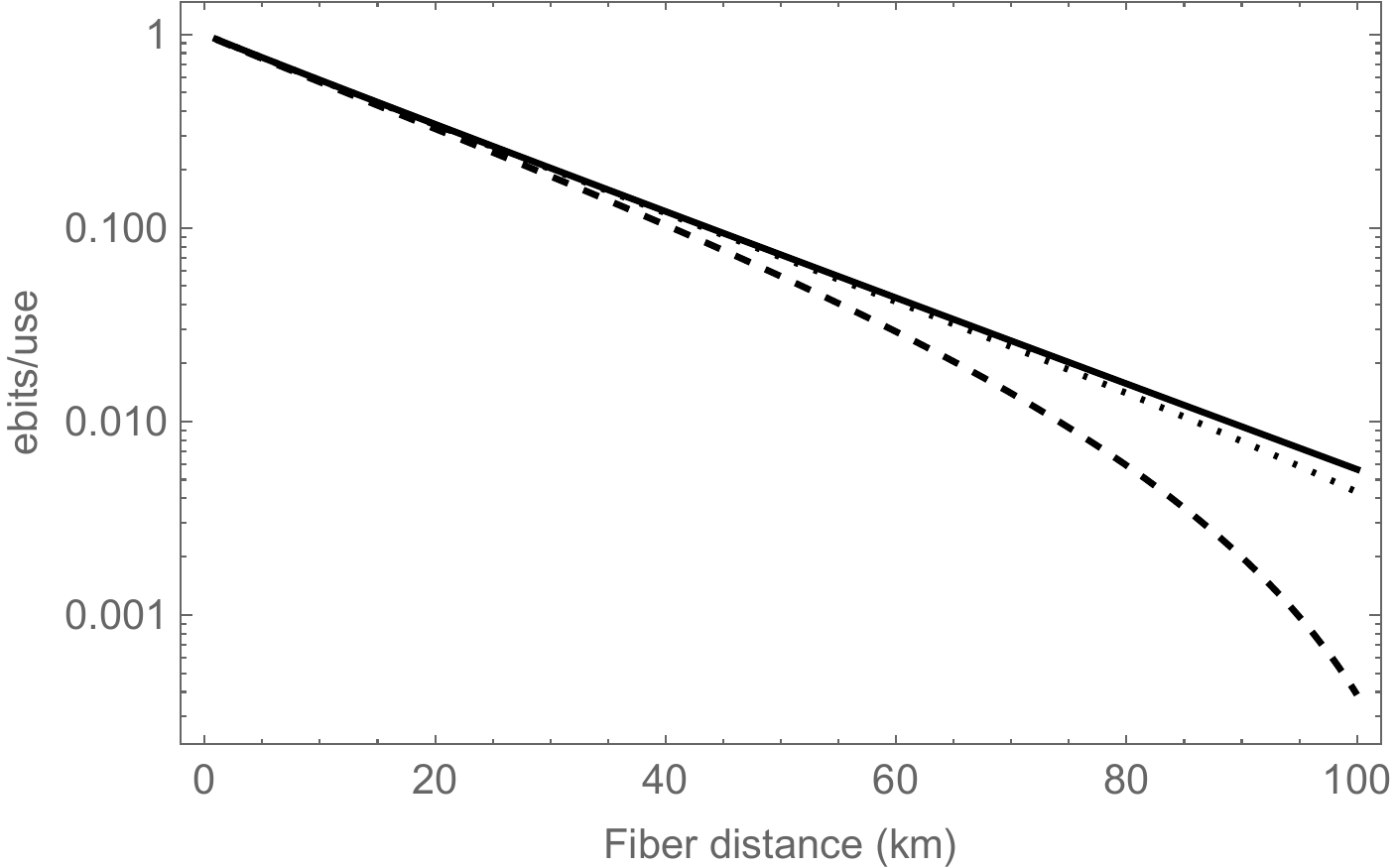}
\caption{We plot the capacity $\mathcal{C}_2(\mathcal{E}_{\mr{EDH}})$ in ebits/use of the erasure-dephasing channel versus fiber distance in km (solid line). $\mathcal{E}_{\mr{EDH}}$ describes entanglement distribution in fiber with active polarization control (dephasing-dominated regime). We also plot an upper bound for $\mathcal{E}_{\mr{EDH}}^{\mr{dc}}$, which accounts for dark counts, with $p_{\mr{dc}}=10^{-2}$ (dashed line) and $p_{\mr{dc}}=10^{-3}$ (dotted line). Fiber parameters are: $\alpha=0.2$~dB/km, $\Delta \nu=100$~GHz and $D_{\mr{PMD}}=0.1$~ps/$\sqrt{\mr{km}}$.}
\label{fig:bounds}
\end{figure}

Given a fiber with transmissivity $\eta(d)$ and PMD probability $p(d)$, we can express the previous bounds in terms of the distance $d$. Here, we are interested in the best performance, so we consider the dephasing-dominated regime. In Fig.~\ref{fig:bounds}, we plot the two-way assisted capacity of the erasure-dephasing channel. For typical parameters, we can see that the distribution of polarization entanglement is indeed feasible with high rates at long distances. For example, with a clock of 1 GHz and the parameters considered in Fig.~\ref{fig:bounds}, we see that Alice and Bob can share about $5 \times 10^6$ ebits per second at 100~km of fiber.

\section{Erasure-Pauli with dark counts}
We can extend the previous model to incorporate receiver-side imperfections, most notably detector dark counts. In this more realistic setting, a no-photon (erasure) event may trigger a spurious detection click, which is then assigned a random polarization outcome. Operationally, this converts part of the erasure component into an effective depolarizing contribution, thereby increasing the overall Pauli error probability.

In the presence of dark counts with probability $p_{\mr{dc}}>0$, the erasure-Pauli channel becomes (see also Fig.~\ref{fig:channel})
\begin{align}
\mathcal{E}_{\mathrm{EP}}^{\mr{dc}}(\rho)
&= 
(1-\eta)\Psi + \eta\mathcal{P}(\rho),\\
\Psi &:=(1-p_{\mr{dc}})|e \rangle \langle e|
+ p_{\mr{dc}}I/2. \label{EPdc}
\end{align}
This channel induces a postselected (click-conditioned) effective qubit channel
\begin{align}
\mathcal{E}_{\mathrm{click}}(\rho)
&=
\frac{\eta\,\mathcal{P}(\rho)
+ (1-\eta)p_{\mr{dc}}I/2}
{\eta + (1-\eta)p_{\mr{dc}}},
\end{align}
which closely resembles the formula of the quantum bit error rate (QBER) in quantum key distribution~\cite{QKDrev1}.

It is important to note that we may rewrite Eq.~\eqref{EPdc} as an equivalent erasure-Pauli cannel as follows
\begin{align}
\mathcal{E}_{\mathrm{EP}}^{\mathrm{dc}}(\rho) = (1-\eta')\ket e\!\bra e + \eta'\mathcal{P}'(\rho),\label{EPdc2}
\end{align}
where we have defined the effective transmissivity
\begin{align}
\eta':= \eta + (1-\eta)p_{\mathrm{dc}} = 1 - (1-\eta)(1-p_{\mathrm{dc}})\label{etap}
\end{align}
and the effective Pauli channel $\mathcal{P}'$ with distribution
\begin{align}
p'_k = r p_k
+ \left(1- r\right)/4,~~~r:=\frac{\eta}{\eta'}. \label{EPnewD}
\end{align}
It is clear that Eqs.~\eqref{EPbounds},~\eqref{FIg} and~\eqref{EPthreshold} are extended to dark counts by replacing $\eta \rightarrow \eta'$ and $\mathbf{p}=\{ p_k \} \rightarrow \mathbf{p}'=\{ p'_k \}$.

In the depolarizing regime, we derive
Eq.~\eqref{EPdc2} where $\mathcal{P}'$ is a depolarizing channel with probability $p' = 1-r(1-p)$,
so we extend Eq.~\eqref{EDP}, and the associated threshold, up to replacing $\eta \rightarrow \eta'$ and $\mathbf{p} \rightarrow \mathbf{p}'$. However, in the dephasing regime, the original dephasing channel is transformed into a Pauli channel $\mathcal{P}'$ with probability distribution as in the form of Eq.~\eqref{EPnewD}. As a result, instead of Eq.~\eqref{EDH}, we apply the more general Eqs.~\eqref{EPbounds},~\eqref{FIg} and~\eqref{EPthreshold}. These equations will govern the performance of the erasure-dephasing channel with dark counts $\mathcal{E}_{\mr{EDH}}^{\mr{dc}}$. 

It is interesting to study the effect that dark counts have on the long-distance entanglement distribution. In Fig.~\ref{fig:bounds}, we see that the upper bound for $\mathcal{E}_{\mr{EDH}}^{\mr{dc}}$ is not far from the capacity in the absence of dark counts. In fact, the deviation becomes evident for longer distances and high values for the dark count probability, such as $10^{-2}$, as in the figure. For smaller values of dark counts, in the order of $10^{-3}$ or less, the deviation becomes negligible.  




\section{Conclusions}
We have introduced the model of erasure-Pauli channel, including some specific forms it may take when the noise has symmetry (depolarizing or dephasing). For this general channel, we have established fundamental bounds for the tasks of entanglement distribution, quantum information transmission, and secret key distribution. We have then applied the model to the distribution of polarization entanglement in optical fiber, discussing two main regimes of operation, depolarizing-dominated and dephasing-dominated.  

\SP{A key novelty of our work is the formulation of a unified information-theoretical model that combines photon loss, polarization decoherence, and detector dark counts within a single quantum-channel description. This enables a capacity-based treatment of realistic polarization-entanglement distribution in optical fiber, going beyond previous analyses that considered these impairments separately or only at the level of detection statistics.}

In the presence of active polarization control, where the PMD regime is dephasing-dominated, we found that high rates can be achieved over long fiber lengths. We also verify that this result is robust to the presence of dark counts in the receiving station by suitably modifying the erasure-Pauli model to account for these extra imperfections. Overall, our results provide an information-theoretical framework to quantify ultimate distance and rate limits for polarization entanglement distribution in optical fiber, clarifying when repeaterless architectures remain viable or need to be assisted by repeaters.

\SP{As future work, it would be interesting to explore the limits of entanglement distribution in free-space paths, where the performance is much more limited due to geometric diffraction, pointing errors, and turbulence.}

\section*{Acknowledgments}
This work was supported by the Integrated Quantum Networks (IQN) Research Hub (EPSRC, Grant No. EP/Z533208/1).


\begin{thebibliography}{99}     

\bibitem {BB84}C. H. Bennett and G. Brassard, ``Quantum cryptography: Public key distribution and coin tossing,'' Proceedings of IEEE International Conference on Computers, Systems and Signal Processing, Bangalore, India, pp. 175–179 (1984).

 \bibitem {Ekert}A. K. Ekert, ``Quantum cryptography based on Bell’s theorem,'' Phys. Rev. Lett. \textbf{67}, 661 (1991).

\bibitem{six} D. Bruss, ``Optimal Eavesdropping in Quantum Cryptography with Six States,'' Phys. Rev. Lett. \textbf{81}, 3018 (1998)

\bibitem{QKDrev0} N. Gisin, G. Ribordy, W. Tittel, and H. Zbinden, ``Quantum cryptography,'' Rev. Mod. Phys. \textbf{74}, 145 (2002).

\bibitem{QKDrev1} S. Pirandola, U. L. Andersen, L. Banchi, M. Berta, D. Bunandar, R. Colbeck, D. Englund, T. Gehring, C. Lupo, C. Ottaviani, et al., ``Advances in quantum cryptography,'' Advances in Optics and Photonics \textbf{12}, 1012 (2020).

\bibitem{sensing1} C.L. Degen, F. Reinhard, and P. Cappellaro, ``Quantum sensing,'' Rev. Mod. Phys. \textbf{89}, 035002 (2017).

\bibitem{sensing2} S. Pirandola,  B. R. Bardhan, T. Gehring, C. Weedbrook, and S. Lloyd, ``Advances in photonic quantum sensing,'' Nature Photon \textbf{12}, 724 (2018). 

\bibitem{Cirac1999}
J.~I. Cirac, A.~K. Ekert, S.~F. Huelga, and C.~Macchiavello, ``Distributed quantum computation over noisy channels,'' Phys. Rev. A \textbf{59}, 4249 (1999).

\bibitem{Kimble2008}
H.~J. Kimble,
``The quantum internet,'' Nature \textbf{453}, 1023 (2008).

\bibitem{QC1} M.~A. Nielsen and I.~L. Chuang, ``Quantum Computation and Quantum Information: 10th Anniversary Edition,'' Cambridge University Press, Cambridge (2010).



\bibitem{Monroe2014}
C.~Monroe, R.~Raussendorf, A.~Ruthven, K.~R. Brown, P.~Maunz, L.-M. Duan, and J.~Kim,
``Large-scale modular quantum-computer architecture with atomic memory and photonic interconnects,''
Phys. Rev. A \textbf{89}, 022317 (2014).

\bibitem{QC2} J. Preskill, ``Quantum Computing in the NISQ era and beyond,'' Quantum \textbf{2}, 79 (2018).

\bibitem{Pirandola2019}
S.~Pirandola,
``End-to-end capacities of a quantum communication network,'' Commun. Phys. \textbf{2}, 51 (2019).


\bibitem{Treiber2009}
A.~Treiber, A.~Poppe, M.~Hentschel, D.~Ferrini,
T.~Lor\"unser, E.~Querasser, T.~Matyus,
H.~H\"ubel, and A.~Zeilinger,
``A fully automated entanglement-based quantum cryptography system for telecom fiber networks,''
New J. Phys. \textbf{11}, 045013 (2009).

\bibitem{Wengerowsky2019}
S.~Wengerowsky, S.~K. Joshi, F.~Steinlechner, J.~R. Zichi,
S.~M. Dobrovolskiy, R.~van~der~Molen, J.~W.~N. Los,
V.~Zwiller, M.~A.~M. Versteegh, A.~Mura, D.~Calonico,
M.~Inguscio, H.~H\"ubel, L.~Bo, T.~Scheidl,
A.~Zeilinger, A.~Xuereb, and R.~Ursin,
``Entanglement distribution over a 96-km-long submarine optical fiber,'' Proc. Natl. Acad. Sci. USA \textbf{116}, 6684 (2019).






\bibitem{Enote} \SP{Note that our model is necessarily more general than the ``dephrasure channel''~\cite{deph1,condsimulation,deph2,deph3,deph4}, which is restricted to
erasure combined with dephasing noise. By contrast, our erasure-Pauli
construction allows for a generic Pauli noise process on the non-erased
polarization component. Moreover, as we discuss afterwards, the model is further extended beyond the
standard erasure setting by including detector dark counts, which transform
some no-photon events into effective noisy polarization outputs.}

\bibitem{deph1} \SP{F. Leditzky, D. Leung, and G. Smith, ``Dephrasure channel and superadditivity of the coherent information,'' Phys. Rev. Lett. \textbf{121}, 160501 (2018).}

\bibitem{condsimulation} S. Pirandola, R. Laurenza, L. Banchi, ``Conditional channel simulation,'' Annals of Physics \textbf{400}, 289 (2019).

\bibitem{deph2} \SP{V. Siddhu, ``Entropic singularities give rise to quantum transmission,'' Nat. Commun \textbf{12}, 5750 (2021).}

\bibitem{deph3} \SP{V. Siddhu and R. B. Griffiths, ``Positivity and Nonadditivity of Quantum Capacities Using Generalized Erasure Channels,'' IEEE Trans. Inf. Theory, \textbf{67}, 4533-4545 (2021).}

\bibitem{deph4} \SP{A. Nema, A. G. Maity, S. Strelchuk, and D. Elkouss, ``Noise is resource-contextual in quantum communication,'' Phys. Rev. Research \textbf{6}, 033089 (2024).}

\bibitem{PooleWagner1986}
C.~D. Poole and R.~E. Wagner,
``Phenomenological approach to polarisation dispersion in long single-mode fibres,'' Electron. Lett. \textbf{22}, 1029 (1986).

\bibitem{WaiMenyuk1996}
P.~K.~A. Wai and C.~R. Menyuk,
``Polarization mode dispersion, decorrelation, and diffusion in optical fibers with randomly varying birefringence,''
J. Lightwave Technol. \textbf{14}, 148 (1996).

\bibitem{AgrawalFOCS}
G.~P. Agrawal, ``Fiber-Optic Communication Systems,'' 4th ed., Wiley, New York (2012).

\bibitem{PLOB} S. Pirandola, R. Laurenza, C. Ottaviani and L. Banchi,
``Fundamental Limits of Repeaterless Quantum Communications,'' Nature Comm. \textbf{8}, 15043 (2017).

\bibitem{2wayCAP} \SP{The two-way assisted capacities represent the optimal rates that are achievable for tasks such as qubit transmission ($Q_2$), entanglement distribution ($D_2$), and key generation ($K_2$). More precisely, these are the maximum rates at which qubits, ebits or key bits can be distributed over a quantum channel, assuming that Alice and Bob do not have pre-shared entanglement but can perform the most general local operations assisted by two-way classical communication. For example, in a QKD scenario, this means that Bob can send classical information to Alice on how to prepare the quantum systems at the input of the channel, and Alice can send classical information to Bob on how to measure them at the output. These exchanges of information are potentially unlimited. The one described represents a repeaterless scenario, i.e., a communication configuration where Alice and Bob are directly connected by the intermediate quantum channel without any third party in the middle. By contrast, in a repeater-based scenario, the communication between Alice and Bob would be mediated by a middle relay or repeater, say Charlie, connected to Alice via channel $\mathcal{E}_1$ and to Bob via channel $\mathcal{E}_2$. In this case, the end-to-end rate between Alice and Bob could exceed the value of a repeaterless rate achieved over a direct quantum channel 
$\mathcal{E}=\mathcal{E}_2 \circ \mathcal{E}_1 $, where Charlie does not intervene.}


\bibitem{REE1} V. Vedral, M. B. Plenio, M. A. Rippin, and P. L. Knight, ``Quantifying Entanglement,'' Phys. Rev. Lett. \textbf{78}, 2275 (1997).


\bibitem{REE2} V. Vedral and M. B. Plenio, ``Entanglement measures and purification procedures,'' Phys. Rev. A. \textbf{57}, 1619 (1998).

\bibitem{SimeoneBook} O. Simeone, ``Classical and Quantum Information Theory: Uncertainty, Information, and Correlation,'' Cambridge, Cambridge University Press (2026). 

\bibitem{LBe-dep} Note that we may also write the lower bound for the erasure-depolarizing channel. This is simply given by the formula $\mathcal{C}_2(\mathcal{E}_{\mr{EPD}}) \ge \eta[1-H_2(3p/4)-\frac{3p}{4}\log_{2}3]$.

\bibitem{Peres} A. Peres,
``Separability Criterion for Density Matrices'', Phys. Rev. Lett. \textbf{77}, 1413 (1996).






\bibitem{PMDnote1}
Write a polarization qubit in the $\{|H\rangle,|V\rangle\}$ basis as
\begin{equation}
\rho =
\begin{pmatrix}
\rho_{HH} & \rho_{HV} \\
\rho_{VH} & \rho_{VV}
\end{pmatrix},
\end{equation}
where $\rho_{HV}=\langle H|\rho|V\rangle$ is the off-diagonal polarization coherence. Under 1st-order PMD, polarization coherence decays as
$\rho_{HV} \rightarrow e^{-d/L}\,\rho_{HV}$. A depolarizing channel acts on the off-diagonal term as
$\rho_{HV} \rightarrow (1-p)\rho_{HV}$, while a dephasing channel acts as
$\rho_{HV} \rightarrow (1-2p)\rho_{HV}$. Equating these coherence-reduction factors to the PMD case above, we find
\begin{equation}
1-p/p_{\infty}=e^{-d/L},
\end{equation}
where $p_{\infty}=1(1/2)$ for depolarizing (dephasing). 


\bibitem{PMD1} S. Massar and S. Popescu, ``Reducing Polarization Mode Dispersion With Controlled Polarization Rotations,'' New J. Phys. \textbf{9}, 158 (2007).

\bibitem{PMDnote2} 
From first-order PMD theory, the mean squared differential group delay (DGD) 
$\langle (\Delta T)^2 \rangle$ scales as $\langle (\Delta T)^2 \rangle = D_{\mathrm{PMD}}^2\, d$
where $d$ is the fiber length~\cite{PooleWagner1986,AgrawalFOCS}. Requiring the rms DGD to be comparable to the photon coherence time $\tau$
leads to $\tau^2 \simeq D_{\mathrm{PMD}}^2\, L_{\mathrm{DH}}$, where $L_{\mathrm{DH}}$ denotes the decoherence length. The additional factor of $2$ appearing in Eq.~\eqref{PMDcoh} depends on the precise
definition of the coherence time and on the chosen visibility threshold. Under an effective exponential approximation, we can write the single-photon coherence visibility as $V \simeq \exp[-\langle (\Delta T)^2 \rangle/2\tau^2]$, which leads to $V \simeq \exp(-d/L_{\mr{DH}})$, where $L_{\mr{DH}}=2 \tau^2/D^2_{\mr{PMD}}$.





\end{thebibliography}
\end{document}